\documentclass[aps,prl,showpacs,preprint,superscriptaddress,floatfix]{revtex4}

\usepackage{graphicx}
\usepackage{dcolumn}
\usepackage{bm}

\begin{document}

\renewcommand{\vec}[1]{\mbox{\boldmath $#1$}}


\title{Molecular spin resonance in the geometrically frustrated MgCr$_{2}$O$_{4}$ magnet by inelastic neutron scattering}


\author{K. Tomiyasu}
\email[Electronic address: ]{tomiyasu@imr.tohoku.ac.jp}
\affiliation{WPI Advanced Institute for Materials Research, Tohoku University, \\
and Institute for Materials Research, Tohoku University, \\
Katahira 2-1-1, Aoba, Sendai 980-8577, Japan}
\author{H. Suzuki}
\altaffiliation[Present address: ]{Advanced Research Laboratory, Hitachi Ltd., \\
7-1-1 Ohmika, Hitachi, Ibaraki 319-1292, Japan}
\affiliation{Department of Applied Physics, School of Science and Engineering, Waseda University, \\
3-4-1 Ohkubo, Shinjuku, Tokyo 169-8555, Japan}
\author{M. Toki}
\affiliation{Photon Factory, Institute of Materials Structure Science, High Energy Accelerator Research Organization, Oho 1-1, Tsukuba, Ibaraki 305-0801, Japan}
\author{S. Itoh}
\affiliation{Neutron Science Laboratory, Institute of Materials Structure Science, High Energy Accelerator Research Organization, Oho 1-1, Tsukuba, Ibaraki 305-0801, Japan}
\author{M. Matsuura}
\affiliation{Neutron Science Laboratory, Institute for Solid State Physics, University of Tokyo, Shirakata 106-1, Tokai, Ibaraki 319-1106, Japan}
\author{N. Aso}
\affiliation{Neutron Science Laboratory, Institute for Solid State Physics, University of Tokyo, Shirakata 106-1, Tokai, Ibaraki 319-1106, Japan}
\author{K. Yamada}
\affiliation{WPI Advanced Institute for Materials Research, Tohoku University, \\
and Institute for Materials Research, Tohoku University, \\
Katahira 2-1-1, Aoba, Sendai 980-8577, Japan}


\date{\today}

\begin{abstract}
We measured two magnetic modes with finite and discrete energies in an antiferromagnetic ordered phase of a geometrically frustrated magnet MgCr$_2$O$_4$ by single-crystal inelastic neutron scattering, and clarified the spatial spin correlations of the two levels: one is an antiferromagnetic hexamer and the other is an antiferromagnetic heptamer. 
Since these correlation types are emblematic of quasielastic scattering with geometric frustration, our results indicate instantaneous suppression of lattice distortion in an ordered phase by spin-lattice coupling, probably also supported by orbital and charge. The common features in the two levels, intermolecular independence and discreteness of energy, suggest that the spin molecules are interpreted as quasiparticles (elementary excitations with energy quantum) of highly frustrated spins, in analogy with the Fermi liquid approximation. 
\end{abstract}

\pacs{75.30.-m, 75.40.Gb, 75.50.Xx, 75.50.-y, 78.70.Nx}

\maketitle

The concept of geometric frustration was first pointed out by Wannier in 1950 for the classic spin systems in the two-dimensional triangular lattice \cite{Wannier_1950}, and has attracted much interest as a paradigm to bring out a novel paramagnetic state. In a geometrically frustrated magnet, the approximation of classic spin with only a freedom of direction can be broken down, because the special atomic arrangement, based on a triangle and a tetrahedron, gives rise to an inherent macroscopic ground-state degeneracy in the classic ground state \cite{Wannier_1950,Anderson_1956}. Therefore, magnetic quasielastic scattering in the paramagnetic phases was intensively studied by neutron scattering in spinel, pyrochlore, kagom$\acute{e}$, triangular systems, and so on. As a result, characteristic spatial correlations of spins were found, such as a small six-spin cluster \cite{Shiga_1993,Ballou_1996,Lee_2002,Chung_2005,Kamazawa_2004,Suzuki_2007}, a large six-spin cluster \cite{Kamazawa_2003}, a seven-spin cluster \cite{Yasui_2002}, spin ice \cite{Kanada_2002,Fennell_2004}, and short-range order with propagation vector(s) \cite{Matsuura_2003,Nakatsuji_2005}. 

\begin{figure}[htbp]
\begin{center}
\includegraphics[width=0.65\linewidth, keepaspectratio]{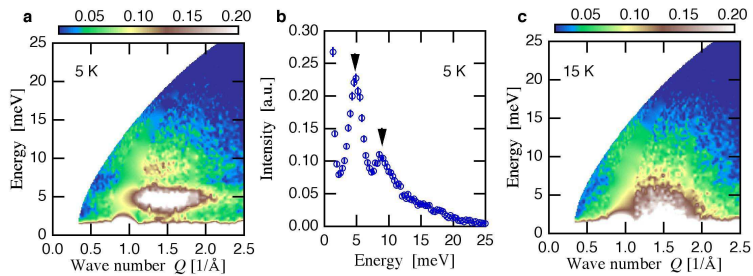}
\end{center}
\caption{\label{fig:pwd} (a)(c) Color images of experimental data of inelastic magnetic scattering in the $Q$ and $E$ space on a powder specimen of MgCr$_2$O$_4$. The color gauges indicate the scattering intensity in arbitrary units. (b) Energy spectra, integrated from 17 to 28 deg in scattering angle, corresponding to the position around Q=1.5 \AA$^{-1}$ in elastic condition. }
\end{figure}

On the other hand, spin fluctuations have been practically unattractive in antiferromagnetic ordered phases in geometrically frustrated magnets so far, because magnetic long-range order appears with lattice distortion, which should eliminate the geometric frustration. However, though for {\it powder} specimens, earlier data of inelastic neutron scattering in the antiferromagnetic phases were reported in the spinels, ZnCr$_2$O$_4$ and MgCr$_2$O$_4$ \cite{Lee_2000,Suzuki_2007}. The $B$ sites in the spinel $A$$B_2$O$_4$ construct corner-sharing tetrahedra, in which a plane of corner-sharing triangles (kagom$\acute{e}$ lattice) and a plane of triangular lattice are alternatively stacked along the [111] direction; an excellent stage for geometric frustration. The two spinels consist of the magnetic ions of Cr$^{3+}$ ($S$=3/2) at the $B$ sites, undergo a first order transition from a cubic paramagnetic phase to a tetragonal antiferromagnetic phase at $T_N$$\simeq$13 K, and exhibit close values of Curie-Weiss temperature, $\theta_W$=$-$370 and $-$390 K \cite{Lee_2000,Klemme_2000,Ehrenberg_2002,Ueda_2006}. The spin systems are almost equivalent in the two chromites. Figures~\ref{fig:pwd}(a), \ref{fig:pwd}(b) and \ref{fig:pwd}(c) show our inelastic neutron scattering data on a powder specimen of MgCr$_2$O$_4$. Since the earlier paper on MgCr$_2$O$_4$ shows only the energy spectrum at $Q$=2 \AA$^{-1}$, we confirmed the overall patterns of $S$($Q$,$E$). The paramagnetic phase exhibits magnetic quasielastic scattering around $Q$=1.5 \AA$^{-1}$ shown in Fig.~\ref{fig:pwd}(c), of which the spatial correlation was identified to be independent antiferromagnetic six-spin clusters (hexamers) by single-crystal specimens \cite{Lee_2002,Suzuki_2007}. But two novel discrete levels are observed around $E$=4.5 and 9.0 meV in the antiferromagnetic phase, as shown in Figs.~\ref{fig:pwd}(a) and \ref{fig:pwd}(b). The first level was proposed to be a localized spin resonance, which cannot be explained by a dimer, and the second level was not mentioned in detail, indicating that the anomalous magnetism of the geometric spin frustration hides even in the antiferromagnetic phase. Therefore, in this paper, we clarify the spatial spin correlations of the two levels in the antiferromagnetic phase by {\it inelastic} neutron scattering on a {\it single-crystal} specimen of MgCr$_2$O$_4$. 

Neutron scattering experiments on a single-crystal specimen were performed on a triple axis spectrometer HER, installed at a cold guide tube of the JRR-3M reactor, and on triple axis spectrometers PONTA and TOPAN, installed at the same reactor, Japan Atomic Energy Agency (JAEA), Tokai, Japan. The energy of the final neutrons was fixed at 3.0, 15, and 14 meV, respectively. A cooled Be filter or a pyrolytic graphite filter efficiently eliminated the higher order contamination. In the HER and PONTA experiments, horizontal focusing analyzers, merging the range of about 7 and 3 deg in scattering angle, were used. The single crystals of MgCr$_2$O$_4$ were grown by the flux method. A specimen of several single crystals with total mass over 300 mg, co-mounted on another triple axis spectrometer AKANE, was enclosed with He exchange gas in an aluminum container, which was set under the cold head of a closed-cycle He refrigerator. 
Neutron scattering experiments on a powder specimen were performed on a direct geometry chopper spectrometer INC, installed at the spallation neutron source, KENS, High Energy Accelerator Research Organization (KEK), Tsukuba, Japan. The energy of the incident neutrons was fixed at 30 meV. A powder specimen of MgCr$_2$O$_4$ with 50 g was synthesized by the solid reaction method, was shaped with an area of 50$\times$50 mm$^2$, and was enclosed and fixed in a similar way.

\begin{figure*}[htbp]
\begin{center}
\includegraphics[width=0.80\linewidth, keepaspectratio]{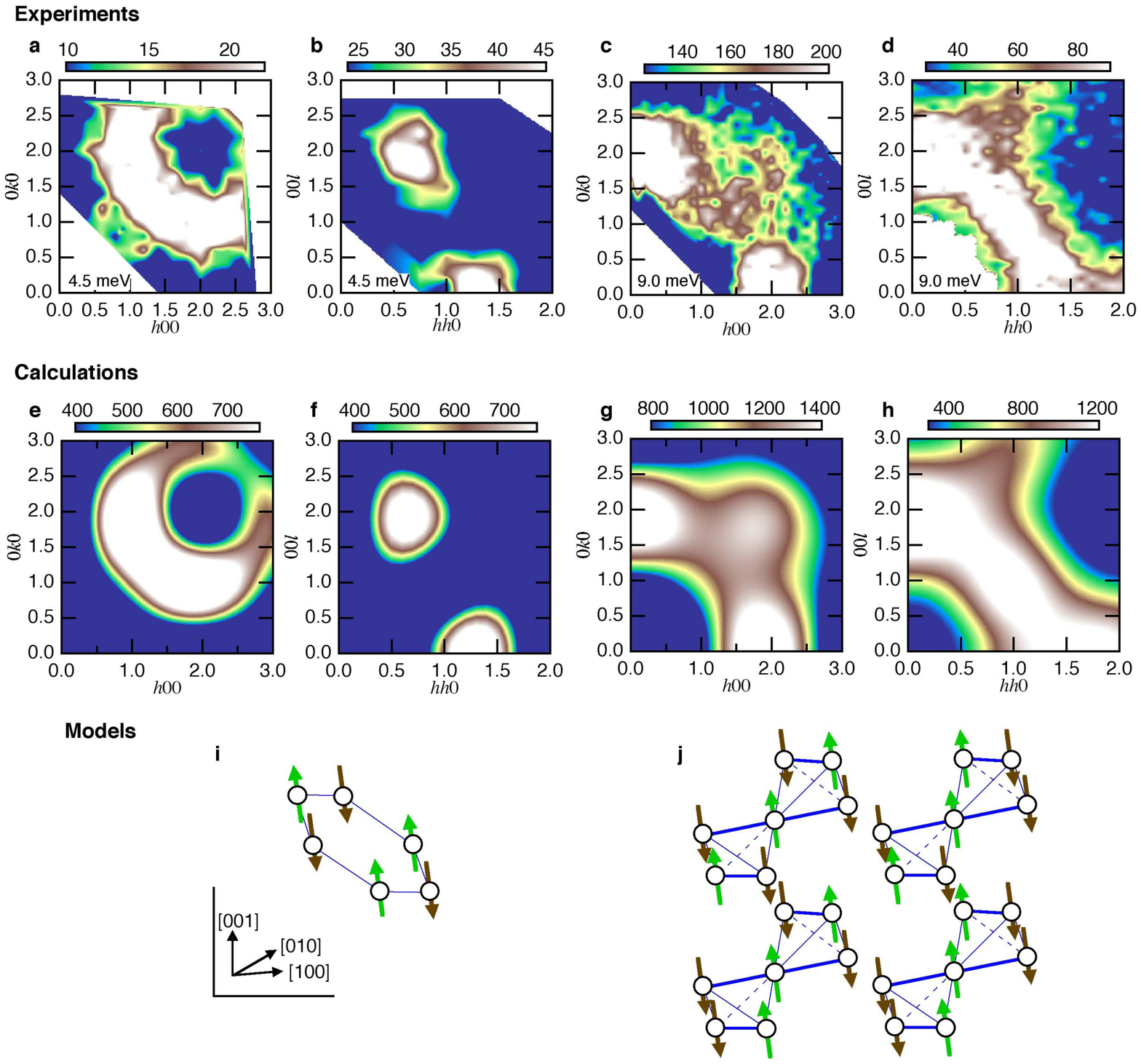}
\end{center}
\caption{\label{fig:sx} Color images of single-crystal inelastic scattering data of MgCr$_2$O$_4$, measured in a constant energy mode. The horizontal gauges indicate the scattering intensity in arbitrary units. (a) and (b) show the maps of experimental intensity, measured at 4.5 meV at 6 K in the $hk0$ zone and the $hhl$ zone on HER, respectively. We also confirmed the same patterns at 1 meV at 15 K. In (a), the area above the [110] line was measured, and that below the line is folded to save beam time. (c) and (d) show the maps, measured at 9.0 meV at 4 K in the $hk0$ zone on PONTA and in the $hhl$ zone on TOPAN, respectively. (e), (f), (g) and (h) depict the calculated patterns of squared form factors of hexamer and heptamer, corresponding to the experimental data (a), (b), (c) and (d), respectively. The schematic pictures of these spin-molecule models are illustrated in (i) and (j). The spins dynamically fluctuate in arbitrary directions with the relative spin correlation shown in (i) and (j). }
\end{figure*}

Figures~\ref{fig:sx}(a) and \ref{fig:sx}(b) are the constant-energy scan data in the $h$$k$0 zone and the $h$$h$$l$ zone, respectively. These data show the first resonance at $E$=4.5 meV and $T$=6 K (antiferromagnetic phase). Surprisingly, the first resonance exhibits the same patterns as the paramagnetic quasielastic scattering in the earlier single-crystal data \cite{Suzuki_2007,Lee_2002}. The magnetic diffuse scattering appears along the boundary of the 220 Brillouin zone in Fig.~\ref{fig:sx}(a) and around 5/4 5/4 0 and 3/4 3/4 2 in Fig.~\ref{fig:sx}(b). 
Such patterns of scattering intensity are reproduced by the model of hexamers in the kagom$\acute{e}$ lattice perpendicular to the [111] direction, shown in Fig.~\ref{fig:sx}(i) \cite{Lee_2002}. The square of form factor of the hexamer is calculated by the relation: 
\begin{equation}
|F(\vec{Q})|^2 = \langle \bigl| \sum_{m=1}^{N} S_{m}\exp({i \vec{Q} \cdot \vec{r}_{m}}) \bigr|^2 \rangle, 
\label{eq:ffsq}
\end{equation}
where $S_m$ is equal to $\pm$1 (relative spin correlation dynamically fluctuating in arbitrary directions), $\vec{r}_{m}$ is a position vector of the spin site $m$, $N$ is the number of vertices of the shape (here 6), and $\langle \phantom{00} \rangle$ means the orientation average over the eight $\langle$111$\rangle$ directions \cite{Lee_2002}. Multiplying the square of the Watson-Freeman's magnetic form factor of Cr$^{3+}$ ion, we obtain the patterns of Figs.~\ref{fig:sx}(e) and \ref{fig:sx}(f) \cite{Watson_1961}, that are in an excellent agreement with those of Figs.~\ref{fig:sx}(a) and \ref{fig:sx}(b). 

Figures~\ref{fig:sx}(c) and \ref{fig:sx}(d) show the data of the second resonance on the $h$$k$0 zone and the $h$$h$$l$ zone, respectively, measured in the constant-energy scan at $E$=9.0 meV. Relatively strong signals are observed around 200 and 020, and a relatively weak one is somewhat distributed around 7/4 7/4 0 in Fig.~\ref{fig:sx}(c). The diluted scattering intensity spreads out so as to connect these three points. In Fig.~\ref{fig:sx}(d), magnetic scattering is relatively strong around 002, and is distributed through 111 towards 3/2 3/2 0. 
The scattering pattern in Fig.~\ref{fig:sx}(d) looks alike that of the independent seven-spin clusters (heptamers), proposed in the pyrochlore material Tb$_2$Ti$_2$O$_7$ with the isomorphic magnetic lattice as the $B$ sites in a spinel material, though only the $hhl$ zone was studied in the Tb compound \cite{Yasui_2002,cmmnt_Tb2Ti2O7_1}. The heptamer consists of two face-to-face triangles in the two kagom$\acute{e}$-lattices and the sandwiched point in the triangular lattice, as shown in Fig.~\ref{fig:sx}(j). We also remark that the same structural unit was recently discussed in the spinel AlV$_2$O$_4$, though the spin correlation is unresolved by neutron scattering \cite{Horibe_2006}. The spin correlation model of Tb$_2$Ti$_2$O$_7$ is explained as follows. All the spins alternatively align on the lines along the [110] and [1$\bar{1}$0] directions, described by the bold blue lines. Fixing the direction of the three spins on the centered line, there are the four types of combinations coming from the remaining two lines, as shown in Fig.~\ref{fig:sx}(j). The square of the form factor of the heptamer is also represented by Eq.~(\ref{eq:ffsq}), where $N$ is replaced by 7, and $\langle \phantom{00} \rangle$ means the average of the 32 types of spin correlations over the eight $\langle$111$\rangle$ directions and the four types (8$\times$4=32). Using the Watson-Freeman's form factor again, we obtain the patterns of Figs.~\ref{fig:sx}(g) and \ref{fig:sx}(h), which are both consistent with those of Figs.~\ref{fig:sx}(c) and \ref{fig:sx}(d). 

%

Thus, the first and second resonance levels in the magnetically ordered phase are identified to be the antiferromagnetic hexamer and heptamer. However, both of the spatial correlations of hexamer and heptamer have been emblematic of the paramagnetic phases of different systems with geometric frustration so far \cite{Lee_2002,Yasui_2002}. This fact suggests the following two points. 
First, a geometrically frustrated magnet essentially implies several magnetic modes. If a geometrically frustrated magnet undergoes a magnetic ordered phase, the first mode in an ordered phase will be the same as in a paramagnetic phase like the hexamer. The candidates of the high-order modes can appear in the paramagnetic phase in other frustrated magnets like the heptamer. 
Second, high symmetry of lattice is {\it dynamically} restored in the magnetically ordered phase with lattice distortion probably by the spin-lattice coupling, often caused by the geometric frustration \cite{Matsuda_2007,Motome_2005,Tchernyshyov_2002}. In addition, as shown in Figs.~\ref{fig:sx}(i) and \ref{fig:sx}(j), all the solid lines connecting the vertices are parallel to the $\langle$110$\rangle$ directions of the $t_{\rm 2g}$ orbitals, $d_{xy}$, $d_{yz}$ and $d_{zx}$, occupied by the three 3$d$ electrons in the Cr$^{3+}$ ion. Therefore, although MgCr$_2$O$_4$ possesses no freedom of orbital and valence, we consider that both spin-lattice coupling and orbital hybridization induce the local itinerancy of 3$d$ electrons (charge freedom) within the molecules, as the mechanism of spin-molecule states. This situation is similar to the quantum resonance in benzene. In fact, benzene is represented by the Kekule structure, in which every vertex possesses three branches (chemical bonds), but the molecular orbital hybridization with non-localized electrons stabilizes the resonance structure of the real benzene, as described by a circle in a hexagon in the chemical structure. 

The present spin-molecule states in the antiferromagnetic phase are experimentally characterized by the intermolecular independence and the discreteness of energy. The two features coincide with the concept of elementary excitation (quasiparticle) in the Landau's Fermi liquid approximation. In the Landau theory, the {\it strongly correlated electrons} behave like independent particles with fundamental energy quanta. Therefore, expanding the approximation to the spin system, we propose that the {\it highly frustrated spins} also act as quasiparticles (geometric quanta or geometrons). With this point of view, the spin correlations in paramagnetic phases are interpreted as the paramagnetic scattering of lowest geometron mode. The modes of geometrons could be classified by the geometric or topologic parameters of spin molecules, such as the numbers of vertices, sides, or holes, as good quantum numbers. 

In summary, the present experiments revealed that the geometric spin frustration sharpens in an antiferromagnetic ordered phase as the component of inelastic scattering. The geometric frustration hides the more profound topics like the geometron picture and the coupling among electronic freedoms and lattice than pure magnetism. Therefore, it is worth studying spin excitations in the antiferromagnetic phase in the higher energy region in many geometrically frustrated magnets. Such investigations are needed to proof the above interpretations and expand the variation of the geometric units.

\begin{acknowledgments}
We thank Professor K. Hirota and Professor K. Iwasa for providing us the extra beam time of the triple axis spectrometers PONTA and TOPAN, Dr. K. Kamazawa, Dr. Y. Yasui, Dr. H. Hiraka and Professor K. Ohoyama for the fruitful discussion, and Mr. T. Asami for the devoted supports of the neutron scattering experiments in ISSP. The authors (K. T., H. S., and M. T.) appreciate Professor K. Kohn and Professor Y. Tsunoda for supporting our study in Waseda University. This work was carried out in the Specially Promoted Research (17001001) in the Ministry of Education, Culture, Sports, Science and Technology of Japan. The work was partially supported by the same Ministry, Grant-in-Aid for Creative Scientific Research (No.16GS0417) and that for Scientific Research (B) (19340090). 
\end{acknowledgments}

\bibliography{MgCr2O4_7_PRL}

\begin{thebibliography}{24}
\bibitem[{cmm()}]{cmmnt_Tb2Ti2O7_1}
\bibinfo{note}{Y. Yasui (private communications). The model of seven-spin
  cluster does not completely reproduce the shape of neutron scattering
  intensity map. Therefore, the spin correlation is also interpreted as the
  short-range order of a complex arrangement like the spin ice in the similar
  materials Ho$_2$Ti$_2$O$_7$ and Dy$_2$Ti$_2$O$_7$, rather than the spin
  molecule.}
\end{thebibliography}

\end{document}